Original Paper

# Investigating the Integrated Digital Interventions Delivered by a Therapeutic Companion Agent for Young Adults with Symptoms of Depression: A Proof-of-Concept Study

Youngjae Yoo[1], PhD, 0000-0001-5412-7769; Minuk Kim[1], BA, 0009-0008-7541-6840; Soyoung Kim[1], BA, 0009-0006-0740-7104; Gayeon Lee[1], BA, 0009-0004-4182-8047; Jinwoo Kim[1, 2, 3*], MA, Prof Dr, PhD, 0000-0002-0162-5446

[1]Graduate Program in Cognitive Science, Yonsei University, Seoul, Republic of Korea
[2]Business Administration, School of Business, Yonsei University, Seoul, Republic of Korea
[3]HAII Corp, Seoul, Republic of Korea

**Corresponding Author:**
Jinwoo Kim, MA, Prof Dr, PhD
Business Administration
School of Business
Yonsei University
Building 212, 5th Fl.
50 Yonsei-ro, Seodaemun-gu
Seoul, 03722
Republic of Korea
Phone: 82 10 6307 2528
Email: jinwoo@haii.io

## *Abstract*

**Background:** Despite the clinical effectiveness of digital interventions for treating depressive disorders in young adults, user engagement and adherence to these programs remain low. To address this issue, the importance of digital therapeutic alliances has been emphasized. Therefore, our study proposes to design a digital intervention that integrates such an alliance with a conversational agent to create a therapeutic companion agent (TCA).
**Objective**: This study aimed to develop a digital intervention prototype using the Wizard-of-Oz method, centered on a TCA that incorporates the following elements of social support: ecological momentary assessment, ecological momentary intervention, behavioral activation, and gamification. The prototype was built using the messaging app KakaoTalk (a mobile social messenger service which is widely used in Korea) and Google Forms (a global online survey tool). We evaluated the clinical efficacy of this intervention at the proof-of-concept level.
**Methods**: Korean young adults aged 20–39 years with mild-to-moderate depressive symptoms, as determined by their nine-item Patient Health Questionnaire (PHQ-9) scores, were recruited online. The intervention group was recruited first and received the digital intervention for 6 weeks. The passive control group was recruited 4 weeks later and instructed to continue their daily routines for 6 weeks without any intervention. The intervention group was guided by a companion agent to perform four daily routine activities for behavioral activation, three daily mood assessments, and meditation. They also received daily reports summarizing their previous day's activities and weekly missions with performance feedback. Both groups were assessed at baseline and at weeks 2, 4, and 6 for depression (Beck Depression Inventory-II), anxiety (Generalized Anxiety Disorder-7), and quality of life (Quality-of-Life Enjoyment and Satisfaction Questionnaire–Short Form).

**Results**: In total, 58 participants were recruited (29 each in the intervention and control groups). One participant from the intervention group dropped out; thus, data from 57 participants were analyzed. The analysis revealed significant differences between the control and intervention groups regarding depression, anxiety, and quality of life at week 6 (P<.05). The intervention group showed a greater reduction in depressive symptoms ($P$=.012, partial $\eta^2$=.066) and a greater improvement in quality of life ($P$<.001, partial $\eta^2$=.139) than the control group over the 6-week study period. Intervention adherence rates were 78% for ecological momentary assessments, 51% for ecological momentary interventions, and 65% for daily routines.
**Conclusions**: The TCA-centered digital intervention was effective in improving participants' depressive symptoms and quality of life. Despite the highly demanding digital interventions, participant dropout was minimal, and adherence was relatively high. Future research should focus on refining the TCA design and conducting more comprehensive evaluations for deeper analysis.

## *Introduction*

**Background**
The increasing global prevalence of depression is a serious issue, emphasizing the need for digital interventions. The global incidence of depression is approximately 23% [1], and it appears to be particularly pronounced among young adults. In the United States, young adults aged 18–24 years have the highest rate of diagnosed depression, at 21.5% [2]; similarly, in several EU countries, the proportion of people aged 18–29 years reporting depression doubled during the pandemic [3]. Similar to the United States and Europe, the prevalence of depression among young adults aged 19–34 years in South Korea has increased by 8.1%, from 18.4% in 2022 to 26.5% in 2024 [4]. Depression in young adulthood is a serious issue as it can disrupt daily functioning [5] and, in severe cases, lead to suicide. Approximately 60% of people who die by suicide had a mood disorder, such as major depressive disorder (MDD) [6].
Despite the seriousness of the problem, traditional face-to-face mental health treatment is often avoided by young people owing to time constraints, financial burdens, and associated stigmas [7]. In a prior study, 54.7% of young adults aged 18–25 years with MDD did not seek any treatment due to cost and other reasons, including a lack of information about mental health services and fear of medication [8]. In this context, digital interventions are gaining attention as effective alternatives for young people, as these interventions can reduce the stigma of mental illness and increase the intention to seek help, thereby improving the awareness and accessibility of mental health care [9]. Consequently, highly accessible and accepted digital interventions are becoming key strategies for addressing depression in young adults.
Owing to the advantages of digital interventions for treating depression in young adults, various methods are being actively researched, focusing on three key factors in their design: 1) selecting core therapeutic techniques, 2) providing just-in-time interventions based on emotional and symptomatic changes, and 3) strategies to increase user engagement. Regarding core therapeutic techniques, behavioral activation (BA), a form of cognitive behavioral therapy (CBT) specifically for depression, has been applied in various digital intervention contexts. BA encourages activities that induce pleasure or a sense of accomplishment to alleviate depressive symptoms [10]. Its clinical validity in digital interventions for depression has been demonstrated in various settings [11-15]. Regarding just-in-time interventions, the design of digital interventions using ecological momentary assessment (EMA) and ecological momentary intervention (EMI) has gained attention. EMA uses mobile devices to repeatedly measure a user's mood and behavior several times a day, allowing for the precise tracking of real-time fluctuations in depressive symptoms [16]. Based on these data, EMI is used to provide personalized interventions at appropriate times, e.g., after several daily surveys or at the user's request [17]. Some studies have also used EMA data analyzed by AI to provide interventions, e.g., emotion recognition questions or messages, to encourage BA [18]. Regarding user engagement, gamification elements are increasingly implemented to encourage the continued use of digital interventions. Previous studies have reported that applying gamification to mental health apps for depression can enhance both user engagement and intervention effectiveness [19]. In a prior study, a group using an app with various gamification elements, including narrative-based quests and feedback-based reward systems, showed improved resilience and a significant reduction in symptoms of depression and anxiety, and these effects reportedly lasted for more than 5 weeks after the intervention had ended [20].

Despite the positive clinical outcomes of these digital interventions, limitations in actual user adherence remain. For example, a 3-year analysis of data from a publicly available digital mental health program showed that only 14.8% of users completed all 12 core modules, whereas 68.6% completed fewer than half, indicating that most users did not complete the entire intervention [21]. Likewise, a meta-analysis reported that less than 50% of participants completed all modules of digital interventions for depression, with an average adherence of 60.7% [22]. For a digital intervention to be effective, sustained use or at least a level consistent with the therapeutic protocol is essential. If adherence is low in a controlled experimental setting, ensuring user engagement in real-world situations may be even more difficult. Considering the low adherence to digital interventions, a sophisticated delivery approach is essential for an effective design, beyond the efficacy of therapeutic techniques, to encourage continuous user participation.

**Digital Therapeutic Alliance (DTA)**

To overcome low engagement and adherence to digital interventions, the importance of a DTA based on emotional connection and trust with the user has been emphasized [23]. A therapeutic alliance, a key factor in the effectiveness of traditional mental health treatments, is defined by three components: the emotional bond between clinician and patient, agreement on therapeutic goals, and collaboration on tasks [24]. These elements are supported by the finding that when trust and empathy are formed between the therapist and patient, the likelihood of treatment continuation and symptom relief increases [25]. A similar DTA concept has been proposed for digital interventions, and research has shown that an emotional bond and trust between the user and intervention system are crucial for increasing intervention persistence and effectiveness [26-28].

In the digital context without face-to-face interactions with a human therapist, the design of conversational agents (CAs) as a digital medium that can provide emotional stability and induce engagement is being discussed as a core element in forming a DTA. In prior research, users felt that anthropomorphized CAs can mimic humans, confirming the possibility of this leading to a better DTA [29]. Additionally, users considered app-based CAs are trustworthy and provide a sense of care [30]. In this context, anthropomorphized CAs can be used as important tools in DTA.

Despite the attention that CAs have received for DTA formation, limitations remain. For example, in "Woebot," a well-known CA for digital intervention, users interacted with the chatbot more than once a day; however, the dropout rate was 17%, indicating limitations in forming a DTA [31]. Another study reported that only 33.6% of users completed all conversations with a topic-based chatbot used in mental health care [32]. These results suggest that existing chatbot-based CAs were not systematically designed to act effectively as deliverers of digital interventions; users did not experience sufficient emotional engagement and motivation, thus failing to form an effective DTA with the CA. This highlights the importance of designing a persona-based CA that can drive user engagement with digital intervention. According to recent research, CA personas play a crucial role in user trust, usability, and engagement [33]. Accordingly, persona-based conversational models significantly improve emotional support and engagement [34]. This suggests that sophisticated persona design for CAs is critical for building an effective DTA for digital interventions.

**Therapeutic Companion Agent (TCA)**

Considering the distinct characteristics of depression treatment among young adults, incorporating the concept of therapeutic companions into interaction design can serve as a valuable approach for forming effective DTAs with CAs. Prior research has defined a companion as a partner who spends time together and fosters an intimate relationship [35, 36]. Social support, particularly within relationships marked by companionship, is a type of social interaction to alleviate problems. Social support is categorized into four types: emotional, appraisal, instrumental, and informational support [37]. When these concepts of companionship and social support are translated into the health domain, therapeutic alliances with companions represent a construct that differs from those with healthcare professionals, e.g. as physicians and counselors. It refers to a peer-like, nonprofessional relationship that enables voluntary self-disclosure, similar to friendship-based interactions [38]. In prior studies on mental health treatment for young adults, non-professional peers or anonymous community members who genuinely care for individuals can be effective sources of emotional support, demonstrating that therapeutic companions are not limited to medical professionals [39].

The concept of a therapeutic companion can also be implemented in digital environments. The formation of human–computer companionships has been a research topic in human-computer interaction (HCI) as digital, virtual, or robotic companions [40-42]. In one study, the perceived human-likeness and consciousness of a chatbot positively impacted the social well-being of users [43]. In another study, a chatbot named Replika was perceived by users as a friend, therapist, and intellectual mirror, acting as a digital companion that provided high-level social support [44]. These digital companions have core characteristics that go beyond simple task performance, including the ability to recognize and emotionally respond to a user's feelings [45], the social interaction capability to form a bond through human interaction [46], and context awareness to perceive and respond appropriately to a user's situation [47].

These characteristics of digital companions are now formalized as TCAs in the digital mental health field. For example, the digital chatbot agent "Wysa" was perceived by users as having a caring presence, and users who interacted with it frequently reported positive effects on their depressive symptoms [48]. Additionally, a conversational interface that analyzed a user's emotional state through real-time conversation and provided feedback and diagnoses has been developed [49], and a chatbot that provided emotional support along with CBT significantly improved mental health indicators such as depression and anxiety [50]. A recent qualitative study using a generative AI chatbot showed that users perceive the chatbot as an emotional sanctuary and a target for emotional connection, suggesting that digital companions can be a valuable emotional resource in therapeutic contexts [51]. Based on these discussions, this study proposes the use of TCA as a delivery medium to increase engagement with and adherence to digital mental health interventions.

### Study Objectives

Therefore, this study aimed to design a TCA-centered digital intervention system for depression in young adults and evaluate its initial effectiveness. We developed a prototype of a digital intervention system that reflects a TCA persona in a Wizard-of-Oz (WoZ) format and evaluated its clinical effectiveness at the proof-of-concept (PoC) level. By empirically demonstrating the usefulness of a TCA design in forming a DTA, this study aims to contribute to improving intervention engagement and adherence, and ultimately treatment effectiveness, for the digital treatment of depression and of various other diseases.

## Methods

### Study Design

This study was designed as a quasi-experimental trial with a nonconcurrent control group. The intervention group was recruited first, whereas the control group was recruited later. Both groups underwent pre- and post-intervention assessments. This design, while potentially lowering result reliability due to confounders from nonrandomization and nonequivalence [52], can be considered a relatively high-reliability quasi-experimental design because both groups were assessed using pre- and post-intervention measures [53]. Furthermore, as this was a PoC study using a prototype rather than a finalized digital system, we could not treat the control group as waitlist control. To prevent participant complaints that might arise from assigning them nonrandomly to different experimental conditions, we recruited the intervention and control groups separately and provided different participation incentives. The clinical effectiveness was evaluated by comparing the intervention and passive control groups.

All methods used in this study were approved by the Institutional Review Board of Yonsei University (7001988-202503-HR-2621-03) and have been conducted in accordance with the World Medical Association's Declaration of Helsinki. The study was also conducted in compliance with the Consolidated Standards of Reporting Trials (CONSORT) guidelines [54]. The participants were fully informed about the study and signed an informed consent form prior to participation. Participants of the intervention group were compensated a maximum of 150,000 KRW based on their participation time and engagement. They received 50,000 KRW for more than 3 weeks, an additional 130,000 KRW for completing the full 6-week period, and an extra 20,000 KRW for participating in the post-study interviews. Control participants received 30,000 KRW for completing the full 6-week period, however, they received no compensation if they dropped out. The participation fees were provided as online vouchers.

### Recruitment

Participants were recruited through convenience sampling from among young adults who reported depressive symptoms. The intervention group was recruited from March 6 to March 23, 2025, whereas the control group was recruited from April 25 to April 30, 2025. Both groups were recruited through online communities and university institutions.

Participants were selected based on the following criteria: (1) Korean young adults aged 19–39 years; (2) depression scores between mild (5 points) and severe (<20 points) based on the nine-item Patient Health Questionnaire (PHQ-9) depression screening tool [55]; and (3) no psychiatric medication for 30 days prior to study application or no changes in their medication.

Participants were excluded from the study if they met any of the following criteria: (1) currently undergoing treatment for a major psychiatric illness requiring hospitalization, such as severe depression, suicidal ideation, bipolar disorder, or schizophrenia and (2) currently receiving CBT for depression, anxiety, or mood disorders or undergoing such treatment in the last 6 months. These criteria were used to identify young adults with mild-to-moderate depression and assess the effectiveness, acceptability, feasibility, and potential concerns of a digital intervention system for depression.

### Interventions

The intervention group was provided with a prototype digital intervention system centered on a TCA, which was designed in the form of a chatbot and delivered via the KakaoTalk chat service for 6 weeks. The TCA design was informed by prior research on companionship and social support and comprised the following key components: companion persona, social support (including informational, emotional, appraisal, and instrumental support), and companionship formation [36].

### Companion Persona

To design a TCA that considers the user characteristics and context of alleviating depression in young adults, we chose a dog persona as the digital intervention delivery agent based on prior research highlighting the mental health benefits and depression-alleviating effects of dogs [56, 57]. Among various breeds, the Golden Retriever was chosen because of its gentle and friendly nature, excellent social skills, and high adaptability to people of various ages and psychological difficulties, making it particularly suitable for psychotherapy [58, 59, 60]. Furthermore, Golden Retrievers are often used as therapy dogs, and interacting with them can provide a sense of psychological stability by reducing stress hormones (cortisol) and increasing happiness hormones (oxytocin) in people with various mental health issues, including depression, post-traumatic stress disorder, and anxiety [61, 62]. Therefore, we developed a dog-like TCA model of a Golden Retriever to provide emotional support and connections to young adults, leveraging the psychological effects of animal-assisted therapy. The TCA was named "Happy" to encourage positive self-affirmation.

To enable "Happy" to send daily messages directly to the participants' chat rooms at a participant-defined time, we used the message scheduling system of the KakaoTalk service. To create a sense of familiarity, we used a picture of "Happy" as the KakaoTalk profile image and used dog sounds like "Woof woof" in the natural language text. We also used nonverbal cues, such as images of "Happy" with a smiling face, crying, or walking, to induce participant engagement and enhance emotional rapport.

**Social Support (1)—Informational Support: EMA**
The first social support provided by the TCA was information support related to depression based on EMA. EMA can repeatedly collect data from participants in real time during their daily lives and dynamically capture changes in their emotions, behaviors, and cognitions [63]. For the effective EMA application in depression, it is essential to set detailed indicators for assessment. Based on previous research, we selected three indicators: (1) rumination, as repetitive negative thoughts are linked to depression [64]; (2) pleasant activity (PA), a behavioral variable because the accomplishment of pleasant activities affects pleasure and mood [65]; and (3) negative affect (NA), as fluctuations in positive and negative emotions are associated with depression [66].

The questions for the three indicators were as follows: (1) NA: "How much are you feeling negative emotions (upset, annoyed, numb, depressed, tense, bored, or anxious) at this moment?," (2) rumination: "Right now, how much do you feel trapped in your thoughts?," and (3) PA: "How much of pleasant activities have you accomplished so far today?". Participants used a visual analog scale to indicate their level of feeling from 0 to 10 points. EMA was delivered three times a day (morning, noon, and evening) at preset times via a Google Forms link in a KakaoTalk chat room. A response was considered valid only if it was completed within 1 h of the scheduled time. The participants individually set their EMA times during the system onboarding process.

**Social Support (2)—Emotional Support: EMI**
The TCA's emotional support consisted of EMI content that directly helps with emotional relief. After completing each EMA, the participants were immediately offered EMI on the next page of the Google Form and decided whether to perform it. The EMI comprised mindfulness-based meditation interventions for depression. To allow participants to choose content based on their preferences, the EMI techniques were curated by experts and included walking meditation, body scan meditation, mindful eating meditation, breathing meditation, and ACT-based rumination journaling [67-71]. Digital intervention content was created using the large language model (LLM)-based generative artificial intelligence (AI) GPT-4o (May 13, 2024, version; OpenAI). The short version of EMI was a text of approximately 1 min and was limited to breathing meditation, rumination journaling, and body scan meditation. The text was displayed directly on Google Forms. The long version was in the YouTube video format. The videos were approximately 5–10 min long, with the walking meditation being relatively longer than the others. The videos were created with GPT-4o, with the generated text being dubbed using AI technology from Naver CLOVA Dubbing. Forty videos were produced, with two versions for each of the five intervention types, each tailored to the participants' characteristics (gender, age, and depression level). The completed videos were then uploaded to YouTube and embedded in Google Forms. The EMI design process is described in Supplementary Figure 1.

When a participant completed the EMI, "Happy" provided a praising message with a smiling image and conducted a satisfaction survey using a 7-point Likert scale. The satisfaction score was shared with the participants in the daily report to help them recognize EMI types with which they were satisfied. If a participant did not perform EMI, a crying image of "Happy" was provided along with encouragement to participate in a future EMI.

**Social Support (3)—Appraisal Support: BA**
Individuals with depression are more likely to experience disruptions in the rhythmicity of their daily patterns [72]. Therefore, this study implemented a daily routine-based BA as appraisal support provided by the TCA, enabling participants to reevaluate and reconstruct their everyday lives. Over the 6-week intervention period, participants engaged in daily routine activities to restore their disrupted lifestyle patterns and maintain a consistent daily routine. To determine the daily routine activities that are most important for individuals with depression, this study referenced prior research [73-74]. We included the following components: 1) washing or showering, 2) writing a daily to-do list, 3) walking a minimum number of steps for physical activity, and 4) journaling.

As BA-based digital interventions significantly reduce depressive symptoms by encouraging users to engage in feasible daily activities [75], we also included BA activities in daily routines. Participants selected 3–6 BA activities from five categories, for a total of 30 activities. The categories were (1) interpersonal relationships, (2) education and career, (3) hobbies and leisure, (4) health (physical and mental) and religion, and (5) daily activities [76-77]. A full list of BA activities is provided in Supplementary Figure 2.

As part of the final routine of the day and the digital intervention, participants wrote via Google Forms a daily diary before going to bed. They checked whether they had washed or showered, entered their step counts for the day, and checked the BA activities they had performed. They could also add additional activities. Then, they selected up to three emotions from a total of 20 using the Positive Affect and Negative Affect Schedule (PANAS) scale to describe their daily feelings [78]. Subsequently, they wrote about an event that influenced their selected emotions and an event they were grateful for. Finally, participants wrote words of praise or encouragement. Through this process, the participants were able to positively reevaluate their daily lives.

Afterward, participants received a personalized daily report based on their intervention participation and diary content. This report was designed to motivate participants and help them understand their own states. The daily report was delivered as an image via the KakaoTalk chat room at 3 p.m. every day from the second day of the intervention until the end of the 6-week intervention. The report included a gratitude journal, a word of praise to promote positive self-perception, and an illustrated diary image generated by ChatGPT-4o based on their diary entries (see prompts for image generation in Supplementary Figure 3). The second page of the report included a "Happy Stamp" for completed activities (washing/showering, walking, to-do list, BA activities, diary, and EMA). It also showed the results of their EMA with color-coded severity (green, yellow, and red) and satisfaction ratings for the EMI activities they completed.

**Social Support (4)—Instrumental Support: Gamification**

To provide instrumental social support, the TCA used a gamification strategy. Unlike other social support, gamification was used to provide a game element that can be experienced by actual participants, to include instrumental support that can be used. According to self-determination theory, gamification plays a key role in stimulating a user's intrinsic motivation and fulfilling the basic psychological needs of autonomy, competence, and relatedness [79]. Gamification elements typically include levels, feedback, points, rewards, badges, and challenges [80-82]. Thus, weekly missions, happy miles, levels, and donation challenges were provided to increase participants' immersion and compliance with each intervention element (Supplementary Figure 4).

The leveling-up system was structured such that participants would either advance to the next level or repeat the current level based on their weekly mission performance, which was assigned every Monday morning. Levels 1–3 focused on quantitative goals, whereas Levels 4–6 required achieving qualitative goals centered on BA activities. Successful mission completion was rewarded with "Happy Mileage," ranging from 10 to 60 points depending on the level. Participants also received one point for each EMA completed within a 1-hour window. The maximum number of Happy Mileage points that could be collected through EMA in a week was 21. If a participant collected 100 or more Happy Mileage points during the 6-week study, the researcher donated 100 KRW per point to a stray dog shelter at the end of the study, and the participant received an official sponsorship certificate with nicknames or real names.

The participants received weekly feedback every Monday morning in the form of image cards that summarized previous mission results and provided guidance for the upcoming mission. Each image card consists of three key elements. First, the card displayed the participant's current level and accumulated Happy Mileage points. The second element featured an illustration of "Happy" engaging in outdoor activities. In the event of mission failure, the illustration depicted "Happy" crying. Finally, a tailored message was presented based on mission success or failure. If the mission was successful, an encouraging message for the next mission was provided; if the mission was not accomplished, a message expressing disappointment while motivating success for the subsequent mission was delivered.

**Companionship Formation: Intervention Onboarding**

As the effectiveness of social support is influenced by companionship, an intervention onboarding process was conducted prior to the main experiment to establish a minimal level of companionship between the participants and the TCA. This onboarding was administered using Google Forms directed at the final registered participants. Unlike conventional "page-turning" style formats, the onboarding was designed as an interactive conversation with "Happy." This design enabled participants to naturally recognize key intervention content and its expected effects while simultaneously forming rapport with "Happy." Participants selected responses based on their individual personalities or attitudes; however, regardless of the chosen answers, the overall content and flow of onboarding remained consistent for all participants. Additionally, "Happy" reinforced crucial information toward the latter part of the onboarding, providing an effective reminder of important points.

Another important goal of onboarding was to help the participants set their own intervention goals. Three types of personalized data were collected. First, the participants set their daily EMA and EMI times in 30-minute increments within the morning (9–12 a.m.), afternoon (1–4 p.m.), and evening (6–9 p.m.) windows. They also decided whether they wanted to receive a reminder message 30 min after the initial notification. Second, they set their daily routine times in 30-minute increments, with the recommended times provided for each activity. The participants also set their own step-count goal with a recommended target to encourage a slightly higher goal of improving depression. Finally, they selected the BA activities that they wanted to perform over 6 weeks. Based on prior research showing that positive self-affirmation can enhance self-esteem and reduce depressive symptoms [83], participants concluded the onboarding process by writing a positive self-affirmation to help themselves (see Supplementary Figure 5 for example screenshots of each digital intervention component).

### Passive Control Group

The control group participants received a text message after registration, instructing them to continue their daily lives for 6 weeks. They were also briefly informed about the online surveys they would complete at baseline and after 2, 4, and 6 weeks. Unlike the intervention group, the control group did not receive any daily intervention messages via KakaoTalk or other digital intervention content. However, they could contact the researchers directly via text messages or phone with any questions during the 6-week study period.

### Measures

Participants completed all clinical and engagement measures online using Google Forms. Clinical measures were collected from both intervention and control groups, whereas engagement measures were collected only from the intervention group.

### Clinical Measures

The primary outcome of this study was a reduction in depressive symptoms in young adults as assessed using the Beck Depression Inventory-II (BDI-II) [84]. The BDI-II is a 21-item self-report scale that quantitatively measures depression severity over the previous 2 weeks. Each item was scored on a 4-point scale (0–3), with a higher total score indicating more severe depressive symptoms. Scores of 0–13 were considered normal, 14–19 mild, 20–28 moderate, and 29–63 severe.

Anxiety was assessed using the Generalized Anxiety Disorder-7 (GAD-7) [85]. Participants answered how often they had been bothered by each of the seven items over the previous 2 weeks on a 4-point scale (0=not at all, 3=nearly every day). Higher scores indicated higher anxiety levels. Scores of 0–4 were considered normal, 5–9 mild, 10–14 moderate, and 15–21 severe.

Changes in the perceived quality of life were assessed using the Quality-of-Life Enjoyment and Satisfaction Questionnaire–Short Form (Q-LES-Q-SF) [86]. The Q-LES-Q-SF measures satisfaction with various areas of life (physical health, mood, work, social relationships, etc.) on a 5-point scale (1=very poor, 5=very good). It consists of 16 items, and the total score is calculated by summing the first 14 items. This provides a quantitative assessment of overall life satisfaction.

### Engagement Measures

We measured EMA, EMI, and daily routine compliance over 6 weeks. EMA compliance was evaluated as the proportion of the total EMA responses completed, including both regular (within 1 h of the scheduled time) and voluntary EMA responses. EMI compliance was calculated as the proportion of EMIs completed out of the total EMA responses. Daily routine compliance was calculated as the percentage of days on which participants completed their self-set routines over time.

## Procedure

Individuals wishing to participate in the digital intervention system accessed Google Forms via a QR code on the recruitment poster to complete the online screening. The screening included four personal information questions for identification, three self-assessment questions for eligibility, and the nine-item PHQ-9. A survey was also conducted to determine the schedule for the participant orientation. For those who understood the online description and consented to the subsequent procedures, an approximately 1-hour online orientation was held via Zoom before official registration. Five sessions were conducted between March 19 and March 23, 2025. During this process, each participant used a pre-assigned "nickname" to ensure anonymity. The orientation explained the overall intervention process, and the participants signed an informed consent form. A pre-intervention survey (T0) was conducted to collect demographic and health information, and the participants completed the BDI-II, GAD-7, and Q-LES-Q-SF questionnaires. The final registered participants completed the onboarding intervention, where they learned key information about the study and entered their personal data. The digital intervention system, named "Journey with Happy," was provided for 42 days (6 weeks), from March 24 to May 4, 2025 (Figure 1). During this period, the system was set to automatically send messages to each participant's individual KakaoTalk chat room at the times they had set during the onboarding process. All messages were delivered by "Happy." The daily message content is shown in Figure 2.

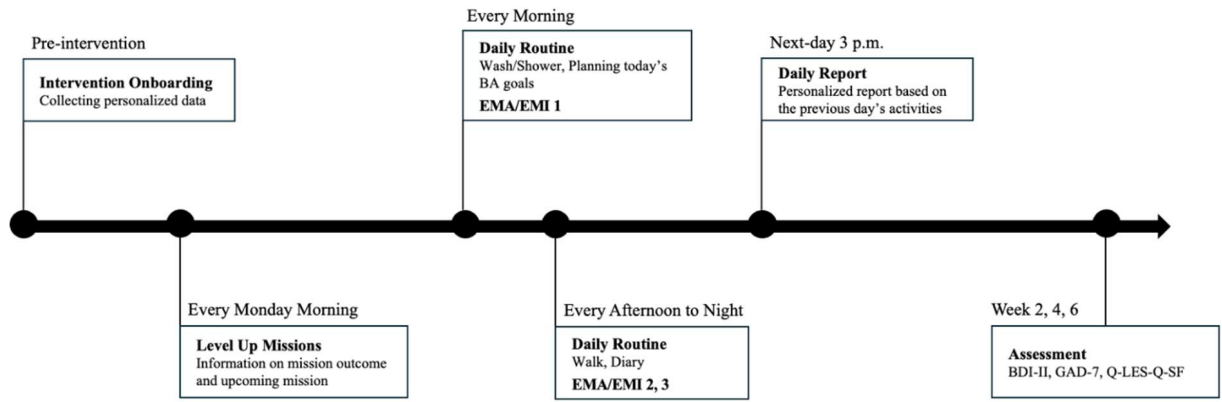

**Figure 1.** Overall intervention structure. BA, behavioral activation; BDI-II, Beck Depression Inventory-II; EMA, ecological momentary assessment; EMI, ecological momentary intervention; GAD-7, Generalized Anxiety Disorder-7; Q-LES-Q-SF, Quality-of-Life Enjoyment and Satisfaction Questionnaire–Short Form.

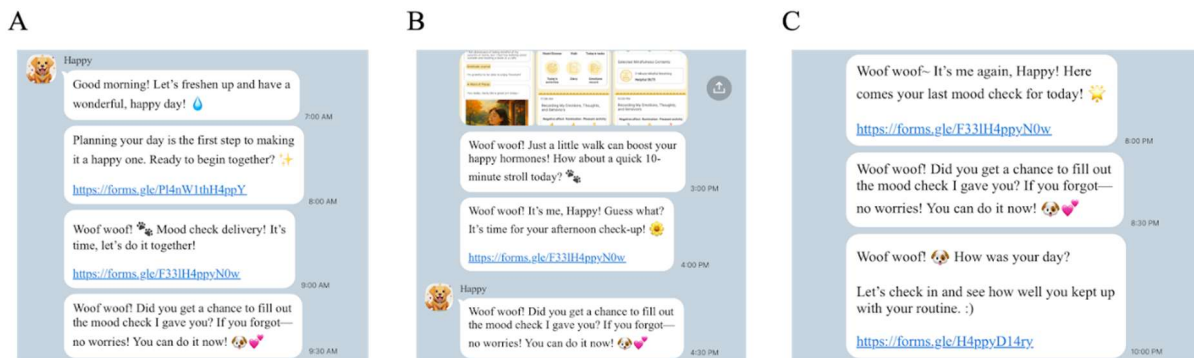

**Figure 2.** Screenshots of KakaoTalk messages. (A) Morning, (B) afternoon, and (C) evening-to-night messages.

First, a morning greeting and a message to wash/shower were sent. Then, at the time selected by the participant for "making today's to-do list," a Google Forms link for "Today's To-Do and Activities" was sent to the individual's KakaoTalk chat room. The participants then listed the BA activities they wanted to perform, wrote a to-do list for the day, and selected one of the BA activities to perform today. At 3 p.m., all participants received a personalized daily report created from their diary entries of the previous day. In the afternoon, the participants received a step-encouragement message at their set time. At the same time, they completed EMA via a Google Forms link for "Recording My Feelings, Thoughts, and Actions" at their set morning, afternoon, and evening times and decided whether to perform EMI after completing EMA. Before bed, they wrote a diary via a Google Forms link for "Looking Back on the Day." The diary form also included sections on any particularly good or bad points of the day. This was included as part of the safety plan to identify any psychological discomfort or negative experiences early and to provide appropriate support.

The intervention group completed online interim surveys (BDI-II, GAD-7, and Q-LES-Q-SF questionnaires) at weeks 2 (T1) and 4 (T2) in addition to the baseline assessment (T0). At the completion of the study, the use of the digital intervention system ended, and a post-intervention survey (T3) was conducted with those who had completed the intervention.

After week 4 of the "Journey with Happy," positive initial reactions to the digital intervention's effectiveness were observed. Therefore, a separate control group was recruited to more precisely compare and verify the relative effect size of the intervention. Control group participants went through the same screening and pre-assessment procedures as the intervention group. Subsequently, from April 28 to June 8, 2025, they completed BDI-II, GAD-7, and Q-LES-Q-SF online surveys at weeks 2 (T1), 4 (T2), and 6 (T3) for 42 days (6 weeks) without receiving any digital interventions or activities.

## Data Analysis

All data analyses were conducted using SPSS Statistics software (version 27.0; IBM Corp). Demographic information was analyzed using independent sample t-tests and chi-square tests. A one-way repeated-measures ANCOVA was used to compare clinical indicators (depression, anxiety, and quality of life) between the control and intervention groups at each timepoint (weeks 0, 2, 4, and 6), with baseline depression and anxiety scores as covariates. A two-way repeated-measures ANCOVA with time (0, 2, 4, and 6 weeks) and experimental conditions (control and intervention) as variables was performed to analyze the differences in the effects on depression, anxiety, and quality of life between the groups, with baseline depression and anxiety scores as covariates. Finally, adherence to EMA, EMI, and daily routine was calculated as a percentage for the intervention group to confirm compliance with the digital intervention program over the entire 6-week period.

# *Results*

## Participants Characteristics

Initially, 29 eligible participants were recruited and assigned to the intervention group. To ensure comparability, an additional 29 participants were recruited and assigned to the passive control group, resulting in a sample size of 58. All participants completed the baseline (T0) assessments, followed by the first follow-up assessment at week 2 (T1). One participant in the intervention group voluntarily withdrew from the study at week 4. As a result, 57 participants (intervention group, n=28; passive control group, n=29) completed additional assessments at weeks 4 (T2) and 6 (T3, endpoint). Figure 3 shows the flow of participants throughout the study.

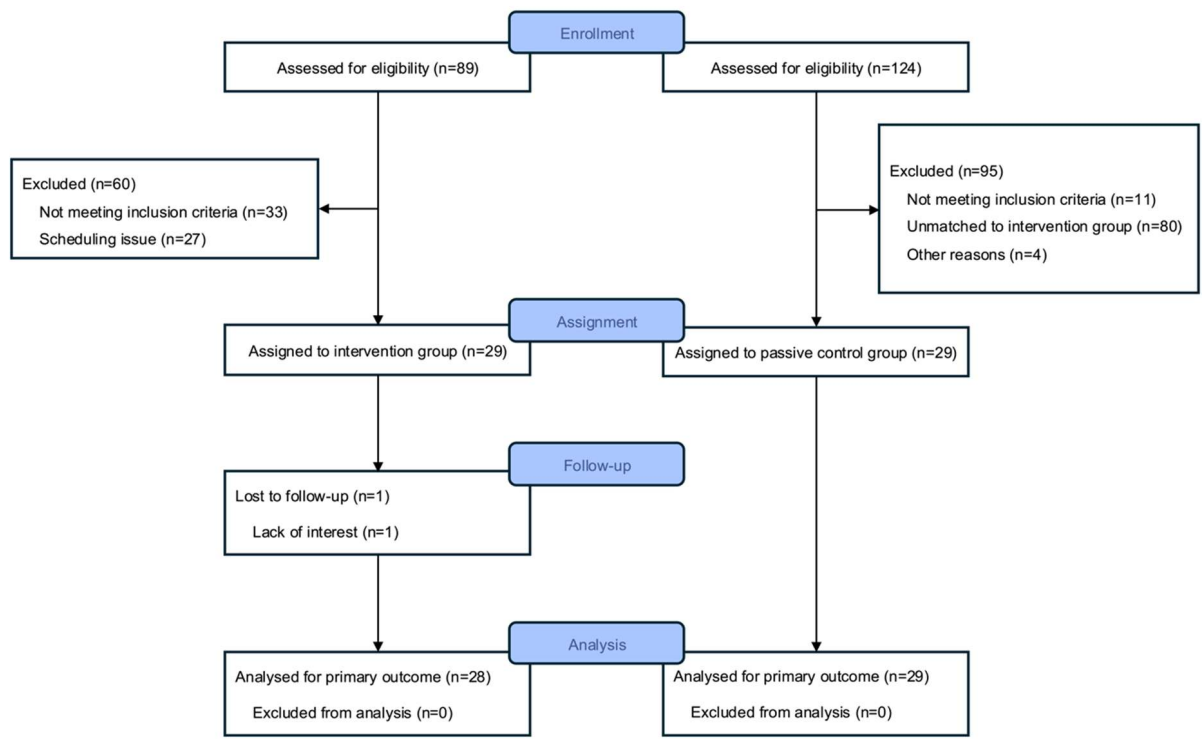

**Figure 3**. Flowchart of participation in the quasi-experiment.

Table 1 shows the demographic and clinical outcomes of all participants. The mean age of the sample was 27.07 (standard deviation [SD]=5.02) years, with 49.1% (n=28) identifying as female. In the intervention group, 71.4% (n=20) were current undergraduate students, and 28.6% (n=8) had graduated from university. In contrast, 24.1% (n=7) of the passive control group were undergraduate students, whereas 75.9% (n=22) had graduated. Only one participant in the intervention group reported being married; similarly, only one participant reported living in a suburban area. In the intervention group, 10.7% (n=3) of participants were diagnosed with depression. Among all participants, only one individual in the intervention group reported using medication (hypnotics). Three participants in the intervention group reported other comorbidities. To verify the homogeneity of the intervention and passive control groups' characteristics, independent t-tests and chi-square tests were conducted. The results indicated significant differences in age and educational status between the intervention and passive control group ($P<.001$).

**Table 1.** Demographics and clinical outcomes in both study groups.

| Demographics and intervention | Intervention (n=28) | Passive control (n=29) | *P*-value |
|---|---|---|---|
| **Age (years), mean (SD)** | 23.71 (4.02) | 30.31 (3.58) | <.001 |
| **Gender (female), n (%)** | 17 (60.7) | 11 (37.9) | .09 |
| **Education, n (%)** | | | <.001 |
|    Currently enrolled in university | 20 (71.4) | 7 (24.1) | |
|    University graduate | 8 (28.6) | 22 (75.9) | |
| **Marital status, n (%)** | | | .31 |
|    Single | 27 (96.4) | 29 (100) | |
|    Married | 1 (3.6) | - | |
| **Residence area, n (%)** | | | .31 |
|    Urban | 27 (96.4) | 29 (100) | |
|    Suburban | 1 (3.6) | | |
| **Psychiatric diagnosis, n (%)** | 3 (10.7) | - | .07 |
| **Current medication use, n (%)** | | | .31 |
|     Hypnotics | 1 (3.6) | - | |
|     None | 27 (96.4) | 29 (100) | |
| **Comorbidities, n (%)** | | | .07 |
|     Gastrointestinal disease | 1 (3.6) | | |
|     Sleep disorder | 1 (3.6) | | |
|     Otitis media | 1 (3.6) | | |
|     None | 25 (89.2) | 29 (100) | |
| **Clinical score, mean (SD)** | | | |
|     (T1) BDI-II[a] | 18.18 (7.20) | 17.86 (8.51) | .45 |
|     (T1) GAD-7[b] | 6.29 (3.43) | 7.17 (4.23) | .42 |
|     (T1) Q-LES-Q-SF[c] | 44.46 (9.28) | 41.07 (6.87) | .07 |
|     (T2) BDI-II | 14.11 (6.06) | 15.24 (11.31) | .60 |
|     (T2) GAD-7 | 5.14 (3.66) | 5.28 (4.42) | .84 |
|     (T2) Q-LES-Q-SF | 47.29 (10.80) | 49.00 (6.78) | .44 |
|     (T3) BDI-II | 11.96 (5.68) | 15.28 (10.20) | .04 |
|     (T3) GAD-7 | 3.89 (3.29) | 4.59 (3.77) | .48 |
|     (T3) Q-LES-Q-SF | 48.25 (10.15) | 47.00 (9.95) | .59 |
|     (T4) BDI-II | 9.50 (6.44) | 13.76 (9.50) | .009 |
|     (T4) GAD-7 | 3.18 (3.45) | 4.97 (3.69) | .03 |
|     (T4) Q-LES-Q-SF | 53.79 (11.06) | 45.93 (9.20) | <.001 |

[a]BDI-II: Beck Depression Inventory-II.
[b]GAD-7: Generalized Anxiety Disorder-7.
[c]Q-LES-Q-SF: Quality-of-Life Enjoyment and Satisfaction Questionnaire–Short Form.
T1, week 0 (baseline); T2, week 2; T3, week 4; T4, week 6 (post-intervention).

**Clinical Outcomes**

The baseline clinical characteristics did not significantly differ between the intervention and passive control groups according to the depression (BDI-II), anxiety (GAD-7), and quality-of-life (Q-LES-Q-SF) scores (all $P>.05$). On average, the participants exhibited mild levels of depressive symptoms at baseline, as reflected by their mean BDI-II scores. Similarly, participants reported mild levels of anxiety according to their GAD-7 scores at the pre-intervention assessment. The mean quality-of-life score, as assessed by the Q-LES-Q-SF (possible range: 14–70), was 39.93 (SD=7.54), suggesting a moderate level of life satisfaction among the participants at study entry. These findings indicate that the two groups were homogeneous in terms of clinical outcomes at baseline, thus supporting the validity of subsequent intervention-related analyses.

Intervention group participants exhibited a steady decrease in depressive symptoms throughout the intervention period. The mean BDI-II score at baseline was 18.18 (SD=7.20), which decreased to 14.11 (SD=6.06) by week 2, 11.96 (SD=5.68) by week 4, reaching 9.50 (SD=6.44) at week 6. In comparison, the passive control group showed relatively stable scores over the same timepoints, with baseline scores of 17.86 (SD=8.51), decreasing slightly to 15.24 (SD=11.31) at week 2, 15.28 (SD=10.20) at week 4, and 13.76 (SD=9.50) at week 6. Between-group differences were not significant at week 2 ($P=.127$) but became significant at week 4 ($P=.038$) and further significant at week 6 ($P=.009$), indicating a meaningful treatment effect favoring the intervention.

To examine changes in depressive symptoms, repeated-measures ANCOVA was conducted with time and condition as fixed factors. The results showed a significant interaction between time and condition ($F(3,165)=3.75$, $P=.01$, partial $\eta^2=.066$), indicating that the pattern of symptom change differed significantly by group (Figure 4). Specifically, participants in the intervention group exhibited a greater reduction in depressive symptoms over time than those in the passive control group. The adjusted mean BDI-II scores were significantly lower in the intervention group (mean difference=–4.246, SE=1.520, $P=.007$), suggesting the intervention's effectiveness in alleviating depressive symptoms.

The observed effect size (partial η²=.066) corresponds to a medium magnitude of effect based on Cohen's conventional benchmarks, in which values of .01, .06, and .14 are interpreted as small, medium, and large, respectively. This finding supports the hypothesis that the intervention elicited clinically meaningful changes in depressive symptoms. Recent research similarly applied and interpreted partial η² values in evaluating digital mental health interventions. For instance, in a randomized study involving college students, Six et al. reported medium-sized effects (e.g., η²=.05) when examining the impact of a conversational mental health app on depressive symptoms, highlighting the practical relevance of such effect sizes in applied digital contexts [87].

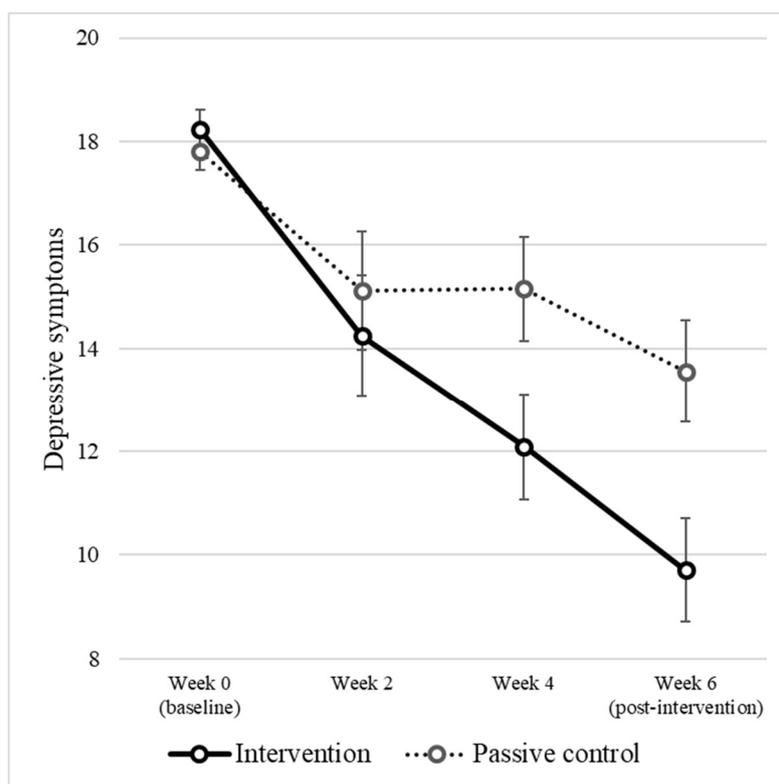

**Figure 4.** Adjusted BDI-II scores over time by group (error bars: 95% confidence intervals). BDI-II, Beck Depression Inventory-II.

Anxiety symptoms in the intervention group progressively declined throughout the study. The mean GAD-7 score at baseline was 6.29 (SD=3.43), which decreased to 5.14 (SD=3.66) by week 2, 3.89 (SD=3.29) by week 4, and further to 3.18 (SD=3.45) by week 6. In contrast, the passive control group started with a baseline score of 7.17 (SD=4.23), which declined to 5.28 (SD=4.42) at week 2 and 4.59 (SD=3.77) at week 4, but slightly increased again to 4.97 (SD=3.69) at week 6. A significant between-group difference was observed at week 6 (*P*=.03), suggesting that participants receiving the intervention experienced lower anxiety levels than the passive control group at the time of post-intervention.

The repeated measures ANCOVA for anxiety symptoms did not reveal a significant interaction between time and condition (*P*=.07), suggesting that the trajectory of change in anxiety symptoms over time did not significantly differ between the groups (Figure 5). However, a significant interaction was found between time and baseline anxiety severity ($F(3,165)=7.67$, *P*<.001, partial η²=.126), indicating that participants' initial anxiety levels moderated their response to the intervention.

The observed effect size (partial η²=.126) is considered to be approaching a large effect. This suggests that anxiety severity at baseline has a meaningful moderating influence on treatment responses. Comparable findings have been reported for recent chatbot-based anxiety interventions. For example, Karkosz et al. demonstrated that a web-based mobile therapy chatbot significantly reduced anxiety symptoms among subclinical young adults, with moderate-to-large effect sizes reported using the GAD-7 as an outcome measure [88]. Their results supported the clinical relevance of digital interventions, particularly in individuals with elevated baseline anxiety.

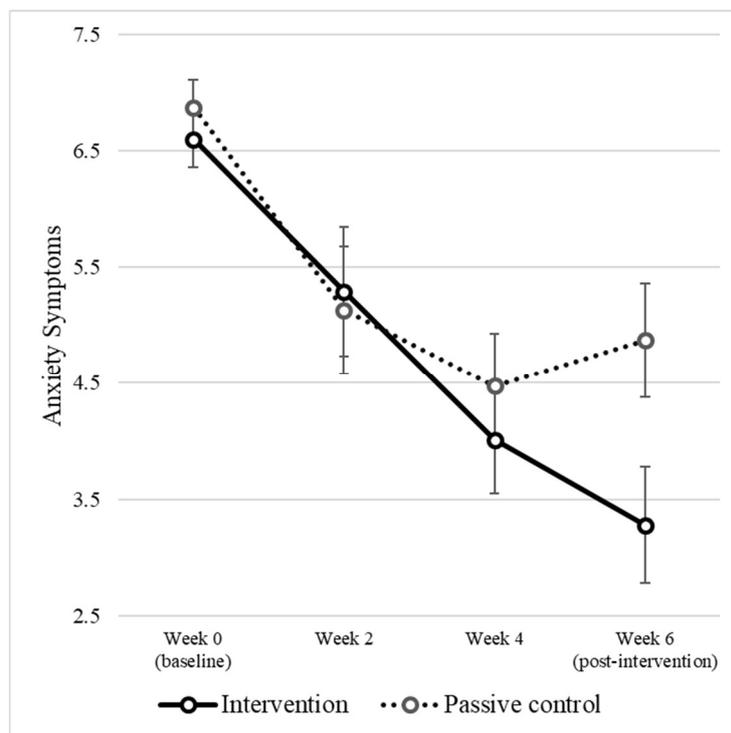

**Figure 5.** Adjusted GAD-7 scores over time by group (error bars: 95% confidence intervals). GAD-7, Generalized Anxiety Disorder-7.

Improvements in quality of life, as measured by the Q-LES-Q-SF, were consistently observed in the intervention group. The mean score at baseline was 44.46 (SD=9.28), which increased to 47.29 (SD=10.80) by week 2, 48.25 (SD=10.15) by week 4, reaching 53.79 (SD=11.06) by week 6. In comparison, the passive control group reported a baseline score of 41.07 (SD=6.87), which increased substantially to 49.00 (SD=6.78) by week 2, declined slightly to 47.00 (SD=9.95) at week 4, and rose again to 49.93 (SD=9.20) at week 6. A significant between-group difference was observed at the endpoint ($P<.001$), indicating that the intervention led to a more robust and sustained improvement in quality of life.

A significant interaction effect between time and condition was observed ($F(3,165)=8.564$, $P<.001$, partial $\eta^2=.139$), indicating that improvements in quality of life varied meaningfully by group (Figure 6). Participants in the intervention group reported significantly higher gains in quality of life (mean difference=−4.467, SE=2.221, $P=.049$) relative to those in the passive control group. The effect size (partial $\eta^2=.139$) approaches the threshold for a large effect, underscoring the clinical relevance of the intervention's impact on life satisfaction.

These findings are consistent with recent evidence from chatbot-based interventions targeting young adults. For example, Karkosz et al. found that a web-based therapy chatbot significantly improved emotional well-being among young adults, and He et al. reported that a mental health chatbot promoted overall psychological health and life satisfaction during the COVID-19 pandemic [88, 89]. Together, these results support the effectiveness of CAs in enhancing quality of life, even in nonclinical populations.

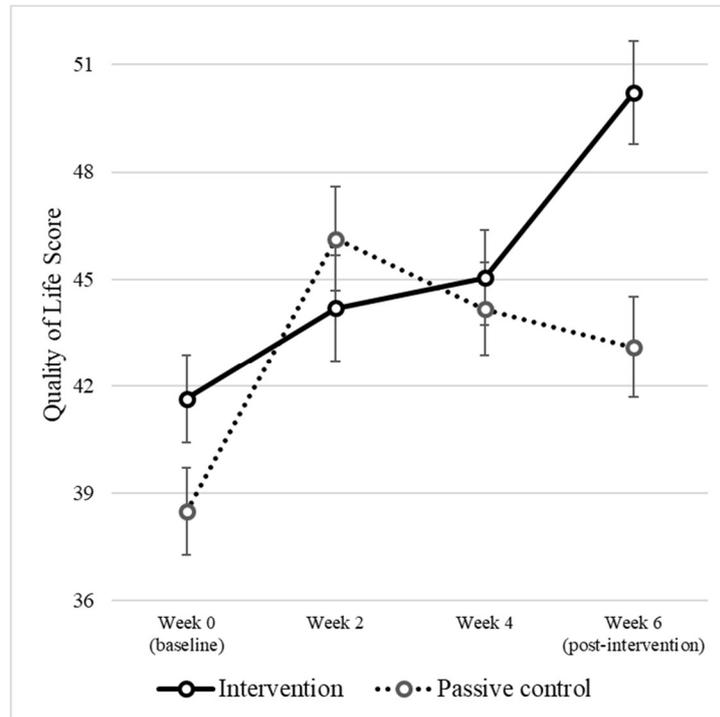

**Figure 6.** Adjusted Q-LES-Q-SF scores over time by group (error bars: 95% confidence intervals). Q-LES-Q-SF, Quality-of-Life Enjoyment and Satisfaction Questionnaire–Short Form.

**Engagement Outcomes**
This study aimed to assess the feasibility of a 6-week intervention by examining participant compliance rates for each component over time (Table 2). Compliance with regular EMA (rEMA), defined as completing a prompted assessment within 1 h, was initially high in week 1 at 82.59% (range: 42.86%–100.00%) but gradually declined to 57.14% (range: 0.00%–90.48%) by week 6. These figures correspond to a range of 58%–96% in week 1 and 31%–82% in week 6, suggesting notable inter-individual variability. This trend is consistent with prior EMA studies that reported rEMA compliance rates typically ranging between 60% and 80% in real-world mobile interventions [90-91].

In addition to completing the rEMAs, the participants were able to voluntarily initiate additional assessments. These unscheduled submissions were aggregated using rEMA to compute the total EMA (tEMA) compliance. The tEMA compliance was calculated as the number of completed EMAs (rEMA + voluntary) divided by the number of scheduled EMA prompts. Notably, tEMA compliance sometimes exceeded 100%, indicating that participants engaged in more assessments than prompted, likely reflecting heightened self-monitoring motivation or the perceived relevance of the assessments. Across the 6-week period, the mean tEMA compliance started at 87.85% (range: 42.86–104.76%) in week 1 and gradually declined to 70.44% (range: 0–104.76%) in week 6. The average compliance across all weeks was 78.46%, which was significantly higher than the scheduled-only rEMA average of 67.46%. This trend echoes findings from prior EMA research suggesting that participants often engage more consistently when given opportunities for spontaneous reporting, potentially because of greater perceived autonomy and contextual salience [92].

EMI compliance, measured as the proportion of EMIs completed following tEMAs, showed a slight upward trend: 49.28% (range: 4.76–137.50%) in week 1, increasing to 53.88% (range: 0.00–100.00%) by week 6. In week 3, some participants completed EMIs at rates as high as 137.50%, surpassing the highest observed tEMA compliance rate (104.76%). This unusually high engagement suggests that certain individuals found the intervention content particularly relevant or personally beneficial, prompting repeated or proactive use beyond what was triggered. Previous research has highlighted that when EMIs are tailored to users' subjective experiences, such as stress, cravings, or mood, they are perceived as more relevant and engaging, which may help sustain higher levels of adherence [91, 93].

Finally, daily routine adherence, defined as the percentage of days participants achieved self-set goals, declined from 75.07% (range: 40.00–94.29%) in week 1 to 59.70% (range: 0–97.14%) by week 6. Although a decline is common in extended ecological designs [90], our overall compliance rates across modalities were comparable to or exceeded those observed in earlier EMA/EMI mental health interventions, underscoring the feasibility of our multicomponent mobile platform.

**Table 2.** Mean compliance rates (%) and ranges over the 6-week study period.

|  | rEMA[a] | tEMA[b] | EMI | Daily routine |
|---|---|---|---|---|
| Week 1 | 82.59 (42.86–100.00) | 87.85 (42.86–104.76) | 49.28 (4.76–104.76) | 75.07 (40.00–94.29) |
| Week 2 | 74.06 (9.52–100.00) | 85.06 (28.57–100.00) | 50.46 (0.00–100.00) | 67.49 (11.43–97.14) |
| Week 3 | 67.32 (0.00–100.00) | 78.33 (4.76–104.76) | 52.7 (0.00–137.50) | 64.93 (14.29–94.29) |
| Week 4 | 63.88 (4.76–95.24) | 76.03 (4.76–104.76) | 51.97 (0.00–110.00) | 59.21 (5.71–100.00) |
| Week 5 | 59.77 (0.00–100.00) | 73.07 (0.00–100.00) | 52.89 (0.00–100.00) | 63.84 (8.57–100.00) |
| Week 6 | 57.14 (0.00–90.48) | 70.44 (0.00–104.76) | 53.88 (0.00–100.00) | 59.70 (0.00–97.14) |
| Total | 67.46 | 78.46 | 51.49 | 65.04 |

[a]rEMA = regular EMA completed by the user at scheduled times.
[b]tEMA = total EMA completed by the user, including both scheduled and user-initiated assessments.
EMA, ecological momentary assessment; EMI, ecological momentary intervention.

## *Discussion*

### **Principal Findings**

This study designed a prototype of a TCA-centered digital intervention delivered through a CA and verified its effects over 6 weeks on depression, anxiety, and quality of life in young adults with mild-to-moderate depression. The results showed a significant interaction effect between time and condition on the alleviation of depressive symptoms and improvement in quality of life in the intervention group compared to the control group. In other words, the intervention group showed a stronger effect on depression and quality of life improvement over time than the control group. However, analyses of changes in anxiety levels did not reveal significant interactions between the control and intervention groups over time. Nevertheless, the anxiety level after 6 weeks was also significantly lower in the intervention group than in the control group. Therefore, despite being designed as a TCA-based digital intervention to specifically target depression, this intervention not only positively affected depression but also alleviated anxiety and improved the quality of life. Importantly, none of the participants in the intervention group dropped out during the 6-week intervention period, and the intervention showed high adherence rates of 78% for EMA, 65% for daily routines, and 51% for EMI. Unlike other interventions, EMI adherence showed an increasing trend over the course of this study. This suggests that the TCA for depression alleviation provided in this study positively impacted the participants' acceptance and engagement with the digital intervention, leading to the observed effects on the alleviation of depression and anxiety and the perceived improvement of quality of life.

The effectiveness of TCA confirmed in this study suggests the importance of a biopsychosocial approach to depression treatment. According to prior research, people with depression prefer treating themselves with medication rather than asking for help from others [94]. Similar to medications, digital interventions for the treatment of depression may be perceived as tools that can be used alone. Contrary to this perception, social support has a positive effect on depression treatment because of the nature of depressive disorders, which often involve a pattern of goal-oriented behavioral failure. For example, social support from significant others, family, and friends was correlated with improved quality of life and reduced depression among college students [95]. A meta-review suggested the effectiveness of peer support by showing that peer support provided by people who have or have had depressive disorders was not statistically different from the effects of CBT-based interventions [96]. Social support can even affect the brain networks of patients with depressive disorders [97]. Considering these findings, a biopsychosocial approach that uses social relationships may be an effective strategy for treating depression. Consequently, the TCA-based digital intervention designed in this study appears to have been accepted by the participants as a social-level treatment beyond the personal level. That is, participants not only experienced digital CBT by using the digital intervention content consisting of BA-based daily routines, EMA, EMI, and gamification themselves but also experienced receiving social support from "Happy," the TCA.

Unlike previous digital intervention designs, this TCA-based digital intervention also appears to have significantly impacted adherence. Typically, existing digital interventions for depression are delivered as digital CBT, in which users experience a pre-developed psychoeducation content series of less than 1 h, with an average of one or two pre-designed sessions per week over several weeks [98, 99], or as digital messaging interventions in the form of short messages delivered several times a day [75, 100]. Unlike existing digital interventions, our study had very high demand, with approximately 10–15 digital messages sent per day, and data entry was required more than five times a day. Despite this demand, only one participant dropped out during the 6-week study, and overall program adherence was comparable to prior studies. This suggests that participants perceived the digital intervention content and messages delivered by the TCA not as a task but as a social interaction with a partner who was trying to help them, as confirmed in previous research using anthropomorphized agents [31, 48]. A previous study investigating the factors determining engagement in digital interventions presented social connectedness as a representative example of how to increase engagement [101]. Therefore, providing a digital TCA-based intervention appears to positively affect acceptance and engagement with the intervention, which is perceived as an interactive experience with the TCA.

**Study Implications**

The TCA use in designing digital interventions for depression, as found in this study, has several important implications. First, it suggests the need for an anthropomorphic concept of intervention using TCA to improve the acceptability and engagement of digital interventions. The TCA developed in this study acted as an anthropomorphic digital intervention concierge, which was found to increase user engagement with digital interventions, thereby improving its clinical effectiveness. The effect of TCAs is expected to be more important in diseases where motivation for performing health behaviors is weak despite the desire to seek treatment, such as depression or substance use disorders, unlike physical diseases such as cardiovascular [102] or musculoskeletal [103] disorders, which show relatively high adherence. Moreover, the TCA-based approach is expected to be less expensive than previously used methods to increase the digital intervention adherence, such as providing financial incentives [104] or serious games [105]. More importantly, the participation levels are expected to increase without a significant increase in fatigue.

Furthermore, this study utilizes various detailed elements in the TCA design, including its core concept, name, appearance, and communication style. Anthropomorphism is a key concept used to consider these detailed design elements. Anthropomorphism is the perception of a nonhuman object's real or imaginary appearance or movement similar to that of a human [106]. For digital interventions and TCAs to be designed effectively, the user must feel that the digital intervention is not a passive system but is being actively delivered by an anthropomorphized agent (i.e., TCA). A prior study that presented a digital assistant design framework to promote physical activity also explained that based on this anthropomorphism concept, an intermediary for delivering digital interventions can be designed by combining three personas (mentor, mate, and trainer) and four avatars (human, older human, dog, and older dog) to interact with the user [107]. This study presented a guideline for anthropomorphic TCA designs, such as the use of the dog-inspired agent "Happy." Based on the study findings, TCA-based designs may increase adherence to and treatment effectiveness of digital interventions for various diseases beyond mental illnesses like depression. Moreover, by using various commercially available generative AI services and applying prompt engineering techniques according to the predefined anthropomorphic style of the TCA, similar to the EMI content created in this study, personalized digital intervention services can be easily concretized [108].

**Study Limitations**

Despite its important implications, this study has several limitations. First, the digital intervention was not a fully automated system; rather, it was provided as a WoZ prototype. The WoZ method has long been used as an effective methodology for evaluating intelligent interfaces based on natural language processing, particularly in the HCI field [109]. However, when applying this method, a risk exists that the delivery of digital interventions may vary depending on the researcher's judgment. Nonetheless, as this study was a semi-automated digital intervention in which the protocol was predefined, this risk was minimized. Furthermore, the WoZ evaluation method is now being actively used in digital intervention studies as well [110, 111]. Considering that this was a PoC study for a nonpatient group before a large-scale clinical trial, the use of the WoZ method was considered sufficiently feasible. Nevertheless, future research will require an evaluation of the clinical effectiveness of the finalized system.

Second, this study did not recruit the control and intervention groups simultaneously and did not use randomization. As the intervention group was recruited first and the control group was recruited later, the participant groups differed in age and education levels. Consequently, confounding variables may have affected the clinical effectiveness of the digital intervention during the 6-week period. Moreover, as randomization and blinding were not used, bias may have occurred in the clinical effectiveness assessment. However, depression, anxiety, and quality-of-life levels did not differ between the control and intervention groups at baseline. All assessments were conducted via online self-surveys, minimizing the possibility of researcher intervention. Nevertheless, follow-up research is needed to conduct a randomized controlled trial with minimal bias through proper sampling, randomization, and blinding.

Third, most participants were not diagnosed with depression. Although applicants with a PHQ-9 score of five or more were included based on recent digital intervention studies on depression in young adults [98, 99], the participants recruited in this study were not diagnosed with depression, so their depression levels were relatively low. Because patients with depression generally have higher PHQ-9 scores [112], the effectiveness of the digital intervention found in this study may not apply to patients with depression. For example, patients with more severe depressive symptoms may not be able to accept the BA techniques used in this study, which might reduce their effectiveness. Future research should include and evaluate a wider range of patients with severe depressive symptoms. In this case, it is mandatory to establish a solid safety plan that includes monitoring by a specialist.

**Conclusion**

Despite some limitations, this PoC study found the potential for a TCA-centered digital intervention that uses a therapeutic companion concept and dog persona to positively impact the alleviation of depression and anxiety symptoms and improve the subjective quality of life in young adults with depressive symptoms. Our study results suggest the possibility of using CAs in future digital interventions for depression in young adults and are expected to effectively increase the acceptability and engagement of digital interventions.

**Acknowledgments**
This article is based on the doctoral dissertation of Youngjae Yoo submitted to the Department of Cognitive Science, Yonsei University. We thank the members of the dissertation committee, Doug Hyun Han, Hyun-seob Cho, Soojin Jun, and Younah Kang, for their valuable feedback.

**Funding Statement**
This work was supported by the Ministry of Trade, Industry & Energy (MOTIE, Republic of Korea) and the Korea Planning & Evaluation Institute of Industrial Technology (KEIT; project number RS-2024-00431485).

**Conflicts of Interest**
None declared.

**Data Availability**
The datasets generated or analyzed during this study are available from the corresponding author on reasonable request.

**Author Contributions**
Conceptualization: YY (lead), SK (supporting)
Data curation: YY (lead), MK (equal), SK (equal), GL (equal)
Formal analysis: YY (lead), MK (supporting)
Funding acquisition: JK (lead), YY (supporting)
Investigation: YY (lead), MK (equal), SK (equal), GL (equal)
Methodology: YY (lead), GL (supporting)
Project administration: JK (lead), YY (equal)
Resources: YY (lead), MK (equal), SK (equal), GL (equal)
Supervision: JK
Validation: JK (lead), YY (equal)
Visualization: YY (lead), MK (equal), GL (equal)
Writing – original draft: YY (lead), MK (supporting), SK (supporting), GL (supporting)
Writing – review & editing: JK (lead), YY (equal)

**Abbreviations**

| | |
|---|---|
| **AI:** | artificial intelligence |
| **BA:** | behavioral activation |
| **BDI-II:** | Beck Depression Inventory-II |
| **CA:** | conversational agent |
| **CBT:** | cognitive behavioral therapy |
| **DTA:** | digital therapeutic alliance |
| **EMA:** | ecological momentary assessment |
| **EMI:** | ecological momentary intervention |
| **GAD-7:** | Generalized Anxiety Disorder-7 |
| **LLM:** | large language model |
| **NA:** | negative affect |
| **PA:** | pleasant activity |
| **PANAS:** | Positive Affect and Negative Affect Schedule |
| **PHQ-9:** | nine-item Patient Health Questionnaire |
| **PoC:** | proof-of-concept |
| **Q-LES-Q-SF:** | Quality-of-Life Enjoyment and Satisfaction Questionnaire–Short Form |
| **rEMA:** | regular ecological momentary assessment |
| **SD:** | standard deviation |
| **TCA:** | therapeutic companion agent |
| **tEMA:** | total ecological momentary assessment |
| **WoZ:** | Wizard-of-Oz |

# Multimedia Appendix 1. EMI Design

1) EMI prompts generated by LLM-based generative AI (ChatGPT-4o)

**Short version:**
I am developing a digital intervention program designed for individuals with depressive symptoms that heightens depressive mood in their daily lives. The target user is [gender], [age] who meets the criteria for [mild/moderate] depression according to the PHQ-9 assessment. The material attached above is a script for [EMI type] alleviating depressive symptoms. Based on the material provided above, please create an intervention text for a one-minute [EMI type] to be delivered to individuals with depression.

**Original Korean Version:**
우울증 환자들이 일상 중 우울감이 높아졌을 때 제공하는 디지털 중재 프로그램을 만드려고 해. 환자는 [성별], [나이], PHQ-9 우울증 평가 기준 [mild/moderate] 수준에 해당해.
위에 첨부한 자료는 우울증 개선을 위한 중재 중 하나인 [EMI 종류]에 대한 스크립트야.
위에 첨부한 자료를 바탕으로, 1 분 동안 [EMI 종류]을 하면서 우울증 환자에게 제공될 디지털 중재용 텍스트를 만들어줘

**Long version:**
I am developing a digital intervention program designed for individuals with depressive symptoms that heightens depressive mood in their daily lives. The target user is [gender], [age] who meets the criteria for [mild/moderate] depression according to the PHQ-9 assessment. The material attached above is a script for [EMI type] alleviates depressive symptoms. Based on the material provided above, please create an intervention text for a five-minute [EMI type] to be delivered to individuals with depression.
Please divide the text into introduction, main body, and conclusion. The introduction, approximately 30 s in length, should present an overview of the program and summarize the evidence regarding the effectiveness of [EMI type] in improving depressive symptoms. The main body should last approximately 4 min and include instructions for the intervention. The conclusion should last approximately 30 s and offer words of encouragement and a closing statement acknowledging the participants' completion of the program.
Please provide only the intervention text as it would be delivered directly to individuals with depression, excluding headings or titles.

**Original Korean Version:**
우울증 환자들이 일상 중 우울감이 높아졌을 때 제공하는 디지털 중재 프로그램을 만드려고 해. 환자는 [성별], [나이], PHQ-9 우울증 평가 기준 [mild/moderate] 수준에 해당해.
위에 첨부한 자료는 우울증 개선을 위한 중재 중 하나인 호흡 명상에 대한 스크립트야.
위에 첨부한 자료를 바탕으로, 5 분 동안 [EMI 종류]을 하면서 우울증 환자에게 제공될 디지털 중재용 텍스트를 만들어줘.
서론, 본론, 결론으로 나눠서 텍스트를 제시해줘. 서론에서는 30 초정도 분량으로 프로그램에 대한 소개와 [EMI 종류]의 우울증 개선에 대한 효과에 대해서 작성해주고,
본론에서는 4 분 정도 분량으로 중재 텍스트에 대한 수행 내용, 결론에서는 30 초 정도 이 프로그램을 잘 마쳤다는 것에 대한 격려와 마무리 멘트를 작성해줘.

그리고, 답변을 할 때, 제목과 같은 텍스트는 제외하고, 우울증 환자들에게 바로 제공될 중재용 텍스트만 만들어줘.

2) EMI Video Edit
To produce the long version of EMI videos, researchers first selected Background Music (BGM) to be used throughout the videos. Researchers searched the term "Meditation BGM Music" on YouTube and sorted videos by view count. The most-viewed videos were selected for each EMI type: body scan meditation and mindful breathing used https://www.youtube.com/watch?v=1ZYbU82GVz4, ACT-based rumination journaling used https://www.youtube.com/watch?v=hlWiI4xVXKY, and walking meditation and mindful eating used https://www.youtube.com/watch?v=3NycM9lYdRI. The videos were edited to ensure that both visuals and audio did not overlap across different EMI types. Personalized intervention scripts were generated for each participant's gender, age, PHQ-9 score, and EMI. These scripts were generated using LLM-based generative AI (ChatGPT-4o). They were then combined with the BGM video using the Naver CLOVA Dubbing service, which enabled control over voice speed and volume through AI-based voice synthesis. The AI-generated narration was customized to ensure a natural and cohesive flow between sentences in alignment with the tone of each EMI.
Finalized videos were uploaded to the researchers' YouTube accounts using the unlisted option.

**Mindful Breathing:**
- Female, 19–24 years, mild: https://youtu.be/-ZCm5J-M_jA
- Female, 19–24 years, moderate: https://youtu.be/qJDCLyd3UIM
- Female, 25–34 years, mild: https://youtu.be/HrutECHaNh0
- Female, 35+ years, moderate: https://youtu.be/PfS-TIlDEwU
- Male, 19–24 years, mild: https://youtu.be/4vXAx1-wvYY
- Male, 19–24 years, moderate: https://youtu.be/PRn6jTZMz4o
- Male, 25–34 years, mild: https://youtu.be/VJyYXHRbeN8
- Male, 25–34 years, moderate: https://youtu.be/CO_X7Zkf8ic

**ACT-based Rumination Journaling:**
- Female, 19–24 years, mild: https://youtu.be/03ILrBxmKSs
- Female, 19–24 years, moderate: https://youtu.be/feDekfNpGy8
- Female, 25–34 years, mild: https://youtu.be/CRFsFp1ygQw
- Female, 35+ years, moderate: https://youtu.be/sMn4lf-HIhA
- Male, 19–24 years, mild: https://youtu.be/aFf0wK1Kxyg
- Male, 19–24 years, moderate: https://youtu.be/sWLQr0DbPLk
- Male, 25–34 years, mild: https://youtu.be/sOCjx-X7dfs
- Male, 25–34 years, moderate: https://youtu.be/ecmVIJrfVpQ

**Walking Meditation:**
- Female, 19–24 years, mild: https://youtu.be/rTeBEECyW2I
- Female, 19–24 years, moderate: https://youtu.be/BeU06WHqCHs
- Female, 25–34 years, mild: https://youtu.be/LMq15_01a_4
- Female, 35+ years, moderate: https://youtu.be/Umn82bPgqTM
- Male, 19–24 years, mild: https://youtu.be/uBwTDgYYxio
- Male, 19–24 years, moderate: https://youtu.be/300RELcbJ5c
- Male, 25–34 years, mild: https://youtu.be/BC2EGqqqo3E
- Male, 25–34 years, moderate: https://youtu.be/PUlOn8-x5FQ

**Mindful Eating:**
- Female, 19–24 years, mild: https://youtu.be/UAfDbPgUP8I
- Female, 19–24 years, moderate: https://youtu.be/woLJFnmV3Vg
- Female, 25–34 years, mild: https://youtu.be/elPSK1jbyLA
- Female, 35+ years, moderate: https://youtu.be/BOyVdMElIcU
- Male, 19–24 years, mild: https://youtu.be/NBLAMMp8cv0
- Male, 19–24 years, moderate: https://youtu.be/qdeM3CmPUY0

- Male, 25–34 years, mild: https://youtu.be/gdb2AT8XTNY
- Male, 25–34 years, moderate: https://youtu.be/MEGqdsdjDeQ

**Body Scan Meditation:**
- Female, 19–24 years, mild: https://youtu.be/5KjW7oj9kvw
- Female, 19–24 years, moderate: https://youtu.be/t1gEd2WHrYA
- Female, 25–34 years, mild: https://youtu.be/H9bxGxjbQV4
- Female, 35+ years, moderate: https://youtu.be/bdZNk_NjcOs
- Male, 19–24 years, mild: https://youtu.be/zG6lxyzMmr8
- Male, 19–24 years, moderate: https://youtu.be/xY3poFcmepk
- Male, 25–34 years, mild: https://youtu.be/WBJZLGCwS-s
- Male, 25–34 years, moderate: https://youtu.be/vE-2TAeYN1I

Note. There was no female participant between 25-34 years with a moderate depression level. Instead, there was one participant who was over 35 years old with a moderate depression level.

# Multimedia Appendix 2. Activities List for Daily Behavioral Activation (BA)

1) Daily routine activities
The study first referred to the *social rhythm metrics* proposed by Szuba et al. [1], specifically the activities of "get out of bed," "start activities," and "physical exercise." Additionally, this study referred to the *self-management strategies for increasing regularity of daily routines during social isolation* presented by Murray et al. [2]. In particular, this study focused on "setting up a routine," such as "getting up at the same time every day," "setting times for a few regular activities each day," and "exercising every day." These activities were examined to inform the determination of daily routine activities in the study.

2) BA activities
This study selected five categories of BA activities based on prior research. The activities proposed in previous publications were modified to align with the objectives and context of this study and are summarized in the table below.

| Category | Original activities in prior studies | Modified activities in the present study |
|---|---|---|
| **Relationships** | Make a special breakfast for my child on Saturday | Make a special meal for my family or friends |
| | Text my friend | Contact a friend |
| | Make special plans with spouse | Make special plans with my family |
| | Tell my child I love them every day | Tell my family I love them or thank you every day |
| | Buy my partner a surprise gift | Buy my family or friends a surprise gift |
| **Education/career** | Learning something new (a language, how to play a musical instrument, etc.) | Learning something new (a language, how to play a musical instrument, etc.) |
| | Read the newspaper every day | Read the newspaper every day |
| | Set a work-related goal | Set a work- or school-related goal |
| | Ask a friend for advice about school | Ask a friend for advice about career |
| | Taking a course on something of interest | Taking a course on something of interest |
| **Recreation/interests** | Go to the park with my son | Go to Hangang park with my family or friends |
| | Playing a musical instrument | Playing a musical instrument |
| | Cooking | Cook my own dish |
| | Listening to music | Singing my favorite music |
| | Photography | Taking photos of my favorite landscapes |
| **Mind/body/spiritual** | Jogging, hiking, walking | Go outside for a walk |

|  | Go to a doctor for a physical/check-up | Go to a doctor for a physical/check-up |
|  | Eat fruit every day | Eat fruit |
|  | Attend a religious service | Attend a religious service |
|  | Baseball or softball | Play my favorite sport (baseball, soccer, etc.) |
| **Daily responsibilities** | Getting my hair cut | Getting my hair cut |
|  | Arrive at work on time | Arrive at work or school on time (not be late) |
|  | Wash my clothes and shoes | Doing laundry or running the washing machine |
|  | Clean the house | Doing housework (washing dishes, etc.) |
|  | Washing the car | Washing the car |

# Multimedia Appendix 3. Image Prompt for LLM-Based Generative AI (ChatGPT-4o)

**English Version:**
Create an image based on the daily diary of a patient diagnosed with depression. The patient is [male/female], [age], and classified as having [mild/moderate] depressive symptoms according to the PHQ-9 criteria. Using the following information provided by the individual, generate a single 1:1 ratio image that reflects a day the patient might perceive as having gone well.
Do not include any text within the image.
Emotion of the day: [participant-selected emotion]
Event that influenced the emotion: [participant-described event]
One thing the participant was grateful for: [participant-described gratitude]
A compliment the participant gave to themselves: [participant's self-affirmation]

**Original Korean Version:**
우울증 환자에게 오늘 하루 동안의 일기를 바탕으로 그림을 만들어주려고 해.

환자는 [여성/남성], [나이], PHQ-9 우울증 평가 기준 [mild/moderate] 수준에 해당해. 환자가 작성한 아래 내용을 참고해서 환자가 하루를 잘 보냈다고 생각할 법한 1:1 비율의 이미지를 한 장 만들어줘.

이미지를 생성할 때, 이미지 안에 글자는 모두 빼줘.

오늘의 감정: [참가자 선택한 감정 입력]

오늘의 감정에 영향을 미친 사건: [참가자 입력한 사건 입력]

오늘의 감사한 일: [참가자 입력한 감사 입력]

스스로에게 칭찬 한 마디: [참가자 입력한 칭찬 입력]

# Multimedia Appendix 4. Gamification Features

**Levels:**
In total, six levels progress from Level 1 to Level 6: Happy Novice, Happy Beginner, Happy Apprentice, Happy Practitioner, Happy Expert, and Happy Master.

**Challenge:**
The missions were designed to be differentiated based on the participants' levels. For Levels 1–3, the focus was on the quantitative completion of missions, encouraging participants to complete as many tasks as possible. By contrast, Levels 4–6 emphasize the qualitative completion of the missions. More complex tasks include at least one BA activity or performing different types of BA activities daily. The missions provided for each level are as follows:
1. Level 0–1: Complete one daily routine every day
2. Level 1–2: Complete two daily routines every day
3. Level 2–3: Complete three daily routines every day
4. Level 3–4: Complete three daily routines every day (including BA)
5. Level 4–5: Complete three daily routines every day (choose a different BA)
6. Level 5–6 (Master Challenge): Complete 100% of the daily routines (choose a different BA)

**Points:**
Happy Mileage is awarded based on the mission completion and level. The Happy Mileage points increased by 10 at each level, allowing participants to earn up to 60 points upon reaching the final level, the Happy Master.
Moreover, Happy Mileage points were provided based on the number of regular EMAs submitted within 1 h of the participant's designated time, ranging from 0 to 3 points per day.

**Reward:**
If the participant accumulated more than 100 Happy Mileage points, the total score was converted to KRW 100 per point. At the end of the study, the total Happy Mileage collected by each participant was calculated in KRW and donated to *Pawinhand*, an animal welfare organization for abandoned dogs. Participants who completed the donation received a sponsorship certificate issued by *Pawinhand* for their real names and nicknames via individual text messages.

**Feedback:**
Every Monday morning, participants received weekly feedback in the form of an image card featuring "Happy." Each image card included the following three elements:
1. The number of Happy Mileage points earned by successfully completing the missions in the previous week.
2. A level-specific illustration of "Happy" engaging in luxurious outdoor activities.
    1. Happy Novice: Relaxing in a park
    2. Happy Beginner: Playing with other dogs at a pet café
    3. Happy Apprentice: Enjoying a swim at a dog swimming pool
    4. Happy Practitioner: Shopping for new toys and treats at a pet shop
    5. Happy Expert: Sunbathing on the beach, wearing sunglasses
    6. Happy Master: Embarking on a luxury cruise trip
3. A personalized motivational message based on the participant's mission results.

# Multimedia Appendix 5. Illustrative Examples of Interventions

(1) Conversational Agent "Happy"
Participants were provided with images of "Happy" expressing various emotions and behaviors, such as walking, smiling, crying, and waving (see below).

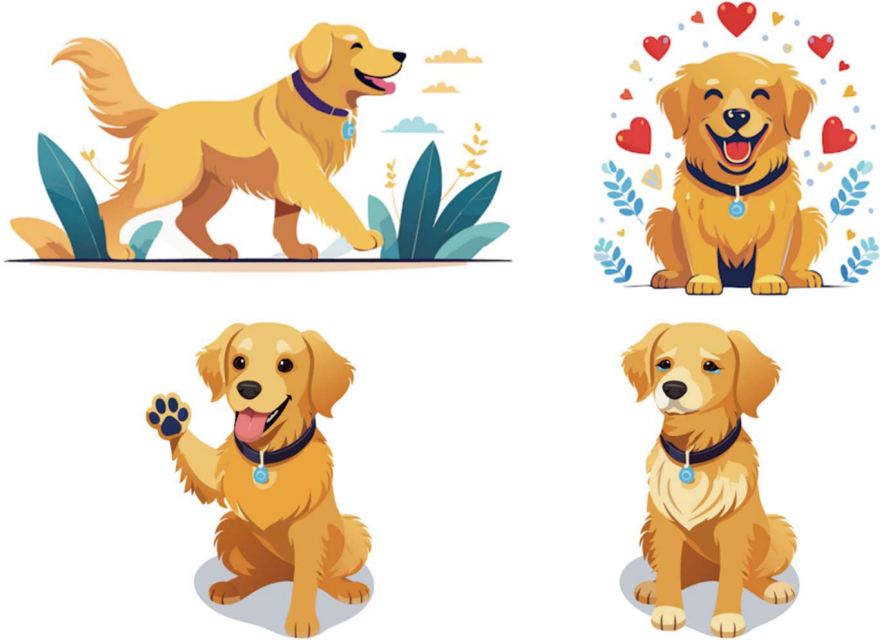

(2) Screenshot of the *Daily Diary and Report* area.
Each night, participants completed a *Daily Diary* (see below). After selecting their emotions, they wrote about the events that influenced their emotions, a gratitude journal, and a self-encouragement message. following day (see below). This *Report* included the *Diary*, an AI-generated image created using an LLM-based generative AI model, proud stamps representing the participants' activities throughout the day, *EMA/EMI* results, and satisfaction levels for *EMI*.

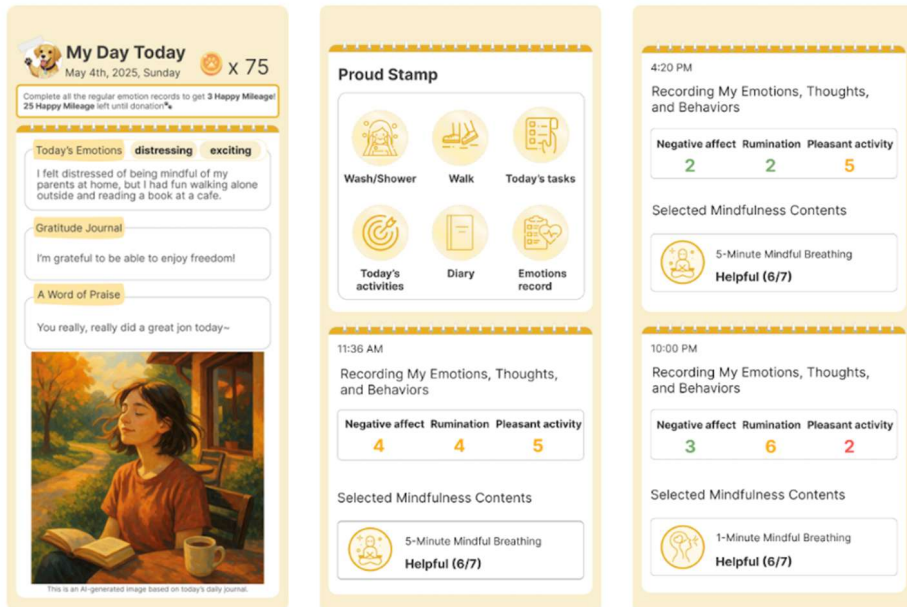

(3) Screenshot of the *EMA/EMI* area.
The types of active *EMA*, negative affect (NA), and pleasant activity (PA) (see below). The participants responded by indicating the position that reflected their experience on a scale of 0 to 10. Types of *EMI*, 1-minute mindful breathing, and 5-minute body scan meditation (see below). The 1-minute *EMI* was presented in text form within Google Forms, whereas the 5-minute *EMI* was linked to a YouTube video.

### Recording Emotions ✎

**At this moment, how strongly are you experiencing any negative emotions (sadness, irritation, insensitivity, glumness, tension, boredom, or anxiety)?**

Please look at the image below and select a number between 0 and 10. 🎯
A score of 0 means you are not experiencing any negative emotions at all, while a score of 10 means you are experiencing negative emotions at their most intense.

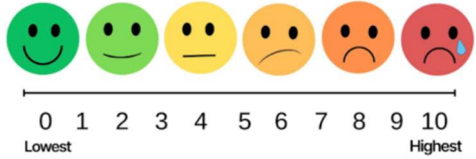

\*

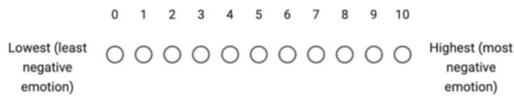

0  1  2  3  4  5  6  7  8  9  10

Lowest (least negative emotion) ○ ○ ○ ○ ○ ○ ○ ○ ○ ○ ○ Highest (most negative emotion)

### Recording Behaviors ✎

**How much have you accomplished pleasant activities so far today?**

Please look at the image below and select a number between 0 and 10. 🎯
A score of 0 means you have not engaged in any pleasant activities, while a score of 10 means you have accomplished the highest level of pleasant activities.

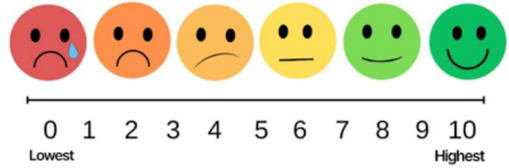

\*

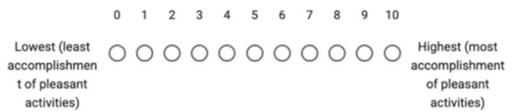

0  1  2  3  4  5  6  7  8  9  10

Lowest (least accomplishment of pleasant activities) ○ ○ ○ ○ ○ ○ ○ ○ ○ ○ ○ Highest (most accomplishment of pleasant activities)

### 1-Minute Mindful Breathing 🙏

Now, begin a 1-minute Mindful Breathing. 🦌

Follow the guide below to begin. :)

It will help calm your mind and bring a sense of ease. 😊

From now on, rest your mind by focusing on your breath for one minute.
Sit comfortably, with your feet flat on the floor and your hands resting on your knees.
You may close your eyes slowly, or keep them open with a relaxed gaze.

Now, breathe in and out slowly, feeling your breath entering through your nose, filling lungs, and leaving smoothly through your mouth.
As you breathe out, just let go of any feelings of sadness or tension you may be experiencing.

Finally, take one last deep breath in and out, and bring your awareness back to the present moment.
Take a moment to appreciate yourself for taking this time to care for your mind and body.

**5-Minute Body Scan Meditation for Relaxation** 🌱🌿

Take a deep breath slowly during a Body Scan Meditation, and notice the sensations in your body from head to toe. 🍃

How about releasing the tension in your body and gradually easing into a deep state of relaxation to make your body comfortable?

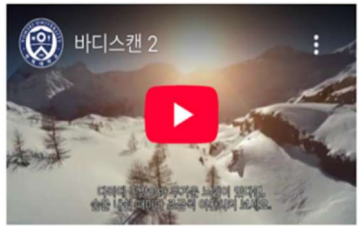

(4) Screenshot of the *Happy Mileage and Level Design* area.
Participants receive feedback on their mission performance from the previous week in the form of an image card featuring "Happy" every Monday morning. This includes the *Happy Mileage*, an illustration of "Happy" engaging in outdoor activities, and an encouraging message. The image on the left represents Happy Apprentice, Level 3. While the image on the right illustrates the state of failure to level up.

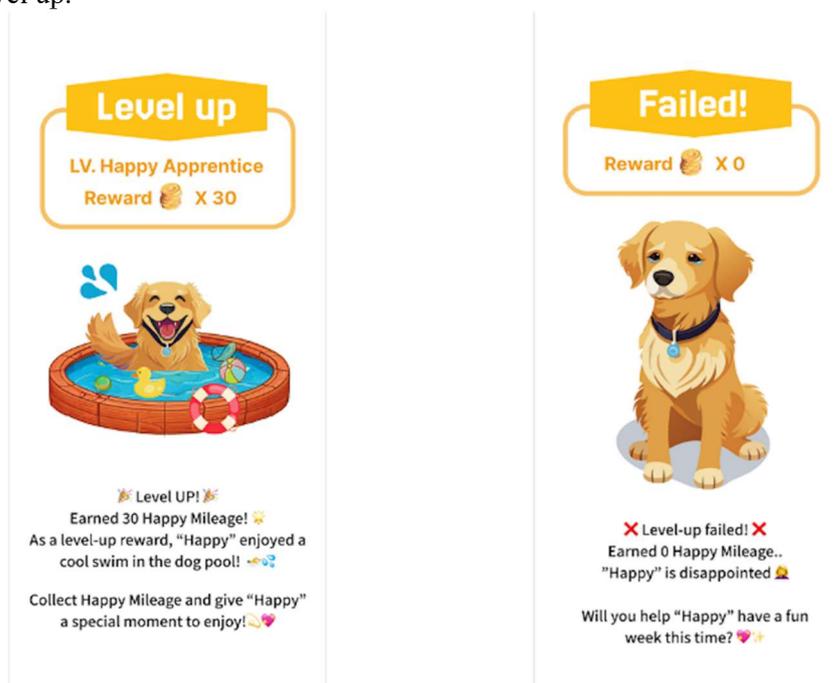

(5) Screenshot of the *Intervention Onboarding* area.
Conversational interaction and rapport building with "Happy" (see left). Personalized data collection for each intervention program (see right). The overall *Intervention Onboarding* process was designed as an interactive conversation with "Happy."

### Recording My Emotions, Thoughts, and Behaviors (3/8)

Of course! You can earn *Happy Mileage* by continuing this journey with *
me.
As your *Mileage* increases, you may feel a little closer to happiness! 😊
The more *Happy Mileage* you collect, the more I feel happy~ 💝
Also, once you reach over 100 *Mileages*, your *Mileage* will be donated to a dog shelter — 100 KRW per *Mileage*! 🐾

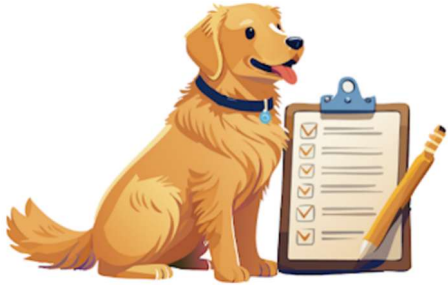

○ Not only me, but "Happy" and other dogs get to feel happy too, right?

○ I see! Can you help me collect as much Happy Mileage as possible?

### Building a Daily Routine (4/6)

When you look over your tasks and make a plan at the start of the day, *
it gives you a sense of satisfaction, as if you've truly owned your day.
And doing this really helps you spend your day more meaningfully! 😊
Please let me know what time of day you would like to write your today's behavioral activation goals.
We recommend doing it before starting your daily activities, ideally **around 10:00 AM** if possible.
Please respond in **30-minute intervals**!⏰ (e.g., 10:00 AM)

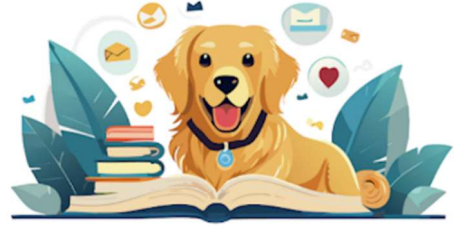

시간

: ▢  AM ▼